# $H^{--}$ events in the ZEUS Detector

E. Accomando[1,2], M. Iori [2], M. Mattioli[2]

[1] Dipartimento di Fisica,
Universita' di Torino, Italy

[2] Dipartimento di Fisica,
Universita' di Roma "La Sapienza", Italy
and
I.N.F.N., sez. di Roma, Italy

April 10, 1995

## Abstract

The supersymmetric left-right models suggest a production of doubly charged Higgs particles. Still there is no evidence for the existence and only some lower limits for their mass are present from the Z decay. Here we investigate the possibility of observing the doubly electronic Higgs decay assuming a $M_{H^{--}} = 50\ Gev$ and $M_{H^{--}} = 100\ Gev$.

# 1 Higgs boson at HERA

The doubly charged Higgs boson, $H^{--}$ are the basic particles of a class of model of elettro-weak interaction beyond the Standard Model with spontaneous parity violation [1,2].

In the recent paper [3] the authors pointed out the possible production of a single double Higgs boson at Hera, in $e-p$ collider at Desy with $\sqrt{s} = 313\ GeV$.

Their calculated the following processes:

$$e^- p \to e^+ p\ H^{--} \quad (1)$$

where

$$H^{--} \to e^- e^-\ (\mu^- \mu^-, \tau^- \tau^-)$$

and

$$e^- p \to \mu^+ p\ H^{--} \quad (2)$$

where

$$H^{--} \to e^- \mu^-\ (e^- \tau^-, \mu^- \tau^-)$$

whose diagrams are shown in Fig. 1

# 2 Study of the topology of the generated events

In order to investigate the possibility of observing the processes (1,2) in the Zeus detector was written a generator based upon the cross section and the decay rate [2]. In this generator the mass of Higgs may be varied also. The decay mode as well the coupling constant $g_{ee}$, $g_{e\mu}$. Fig. 2 shows a plot of the calculated cross section as function of Higgs mass. The dashed line refers to the process (1) and the full line to the process (2).

The first step in the analysis was to investigate the topology of the generated events in order to estimate the fraction for which the lepton would visible in the detector.

As shown before the Higgs boson can decay in two identical lepton ($e^- e^-$, $\mu^- \mu^-$) or in electron- muon ($e^- \mu^-$). In the first case according to the leptonic number and charge conservation in the final state is present a positron in the second case a positive muon. In both situations they follow the electron beam direction (backward respect to the proton beam direction).

The processes (1,2) have the caratterisctics to be diffractive ( high $Q^2$) then the proton goes in the beam pipe (forward direction) and no information in the hadronic calorimeter are left.

## 2.1 Study of reaction $e^-\ p \to p\ e^+\ H^{--} \to p\ e^+\ e^- e^-$

The Zeus Monte Carlo provides a detailed simulation of the detector geometry and resolution based upon the Geant package. The output is read in Zephyr and recontruction of events is carried out.

Using this software in the offline enviroment we have investigated the global properties of the event. For the reason the calorimeter is hermetic we devoted our study to the $e^- - e^-$ Higgs decays in particular to the following reaction

$$e^- \; p \to p \; e^+ \; H^{--}$$

where

$$H^{--} \to e^- e^-$$

As shown in Ref [3] those results exclude an Higgs-boson mass $6.5 < M_{H^{--}} < 36.5 \; GeV$ as function of the coupling strength of the Higgs boson to lepton pairs,$g_{ll}$. We have chosen a Higgs mass value of 50 GeV and of 100 GeV with $g_{ee} = 0.5 \; 10^{-4}$ and $g_{ee} = 0.1$ respectively. An electromagnetic cluster finding algorithm for Uranium calorimeter [6]and CTD finding software [7] was used to identified a global track.

Figs 3a-b show the energy deposited as electromagnetic cluster in the calorimeter and the distribution of the polar angle, $\Theta_l$, of one of the negative leptons in the final state for $M_{H^{--}} = 50 \; GeV$ respectively.

We find the most of the energy relased by the electron in the calorimeter is greater than 20 GeV and the energy distribution of the electromagnetic cluster (Fig. 3a) is peaked at 25 GeV. In Fig 3b we find a deleption of events at small $\Theta_l$ angle due to calorimeter acceptance in forward region.

Figs 4a-b show the same distribution as Fig 3a-b using $M_H = 100 \; GeV$. We find again an excess of events with very low energy deposited in em-calorimeter and the $\Theta_l$ distribution is shifted to low values as expected. Each time the decay track loses the energy deposited in the calorimeter the track can be identified by the cone algorithm as pion or muon.

After a generation of 1000 events ,using the setup of 1994 data taking, with $M_{H^{--}} = 100 \; GeV$ ,by the cluster identification, we find the 88% ± 3% are identified electron , 8% ± 1 pion and the rest are muons and jets. The misindentification of the cluster as pion increases to 9.5% ± 1 when we assume $M_{H^{--}} = 50 \; GeV$.

Fig 5a-b show the $\Delta\phi_{ee}$ angle distribution respectively obtained using only the two highest electromagnetic clusters found in the calorimeter with $E_{cal} > 20 \; GeV$ at least and the Higgs mass reconstructed imposing $150 < \Delta\phi_{ee} < 200$, $\Theta_l > 30$ degrees for the decay leptons. A clear peack at $M_H = 50 \; GeV$ is found. Fig 6a-b show the some distribution of Fig 5a-b assuming $M_{H^{--}} = 100 \; GeV$.

The topology of this sample of events is characterized by a two isolated tracks in the Central Tracking Detector (CTD), back-to-back in the transverse plane leaving large amount of energy in the electromagnetic calorimeter.

The general conclusion to be drawn from these investigations is that the detector acceptance is able to detect two leptons decaying from Higgs boson and is approximately constant if we assume an Higgs mass of 50 GeV and 100 GeV but we lose electron identified track by the calorimeter at small $\Theta_l$ values ($15 < \Theta_l < 40$)

# 3 Background

The nature of the Higgs boson signal chosen with two high energy electron seen in the detector makes the physics background rare. The transverse energy, $E_t$ and longitudinal energy $E_z$ for the decay electrons is plotted in Fig 7-8 rispectively for $M_{H--} = 50\ GeV$ and $M_{H--} = 100\ GeV$ and $E_e > 20\ GeV$. We find the $E_t$ distribution is peacked at 20 GeV and 45 GeV for $M_{H--} = 50\ GeV$ and $M_{H--} = 100\ GeV$ respectively. Appling a cut in tranverse energy, $E_t > 15\ GeV$ we remove possible contribution from elastic $J/\Psi$ where a charge misidentification is present.

# 4 Conclusions

We have investigated one of the production processes of Higgs boson at Hera energies in the framework left-right symmetric model proposed by Senjanovic and Mohapatra. The presence of $M_{H--} > 40\ GeV$ is allowed both theoretically and experimentally assuming $g_{ee} > 10^{-3}$. A remarkable signature enables us to search for the Higgs mass. According with this analysis after 100 $pb^{-1}$ we extimate have $18 \pm 1$ events from Higgs with $M_{H--} = 50\ GeV$ characterized by a two isolated tracks in the Central Tracking Detector (CTD), back-to-back in the transverse plane leaving $E_{Cal} > 20\ GeV$

# 5 Acknowledgments

We thank to Prof. S. Petrarca for valuable discussions.

# 7 Figures

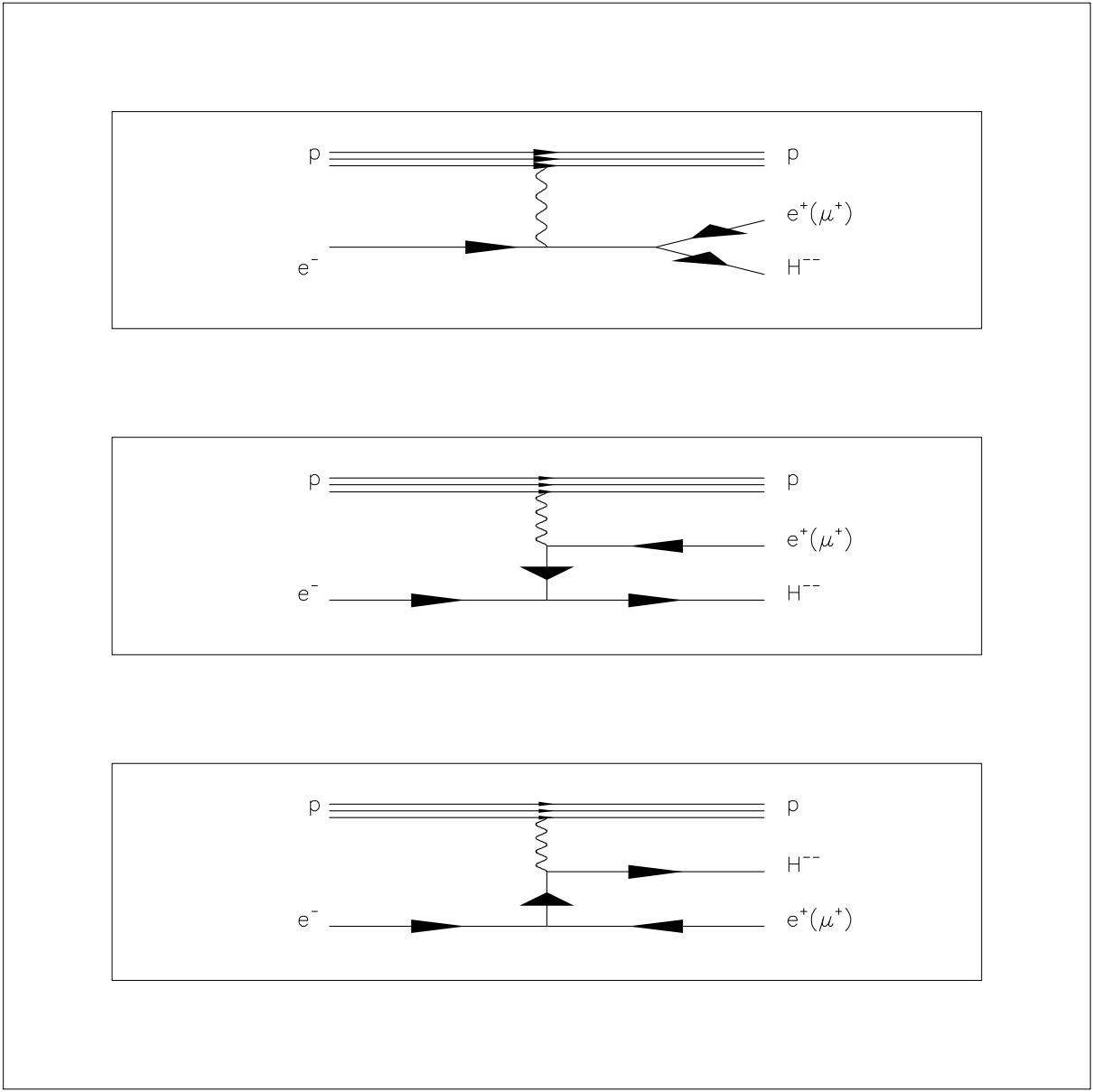

Figure 1: Feynman graphs for single $H^{--}$ production at Hera

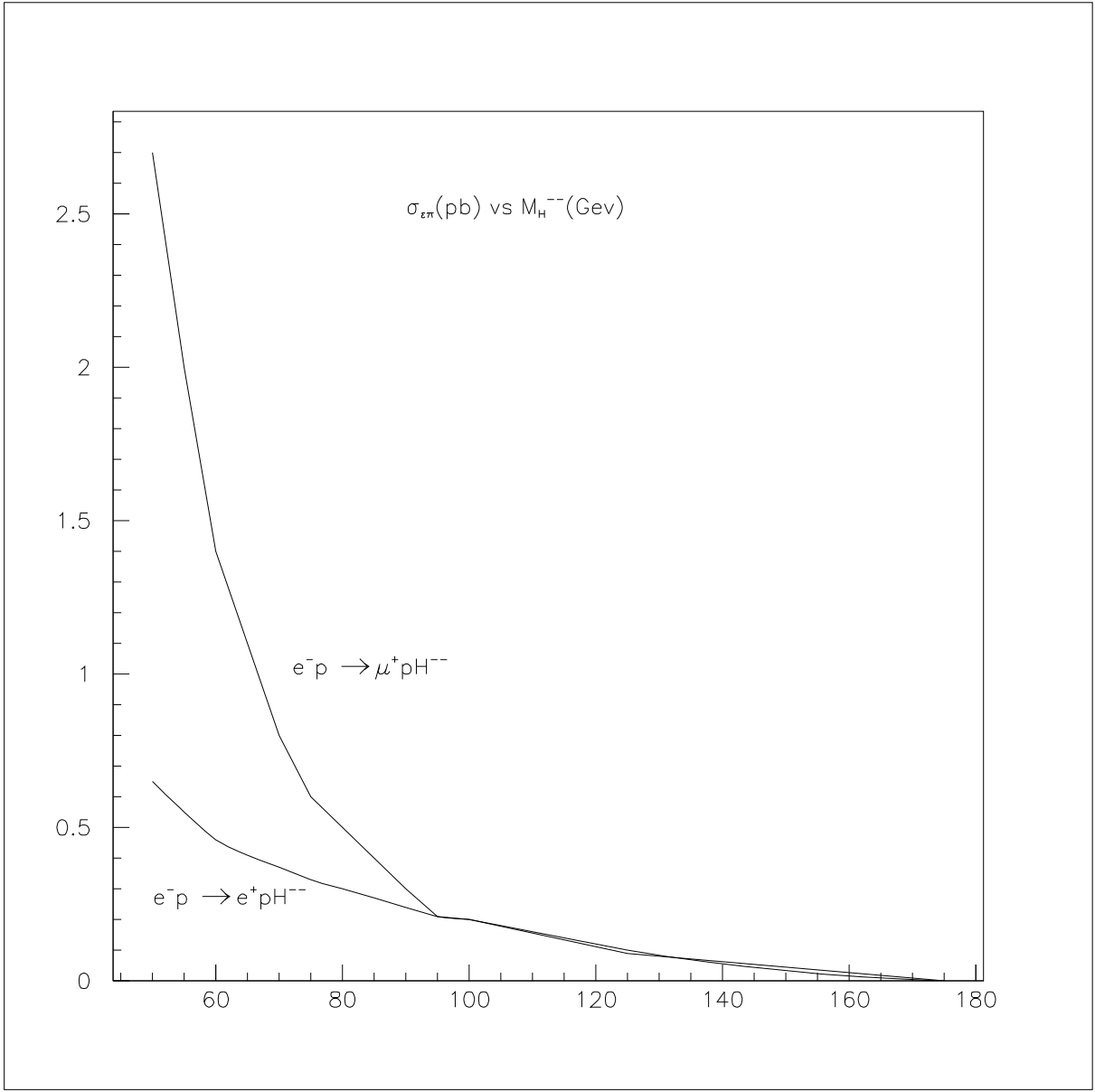

Figure 2: Plot of the total cross sction at $\sqrt{s} = 313\ GeV$ versus $M_{H--}$ from Ref [3]

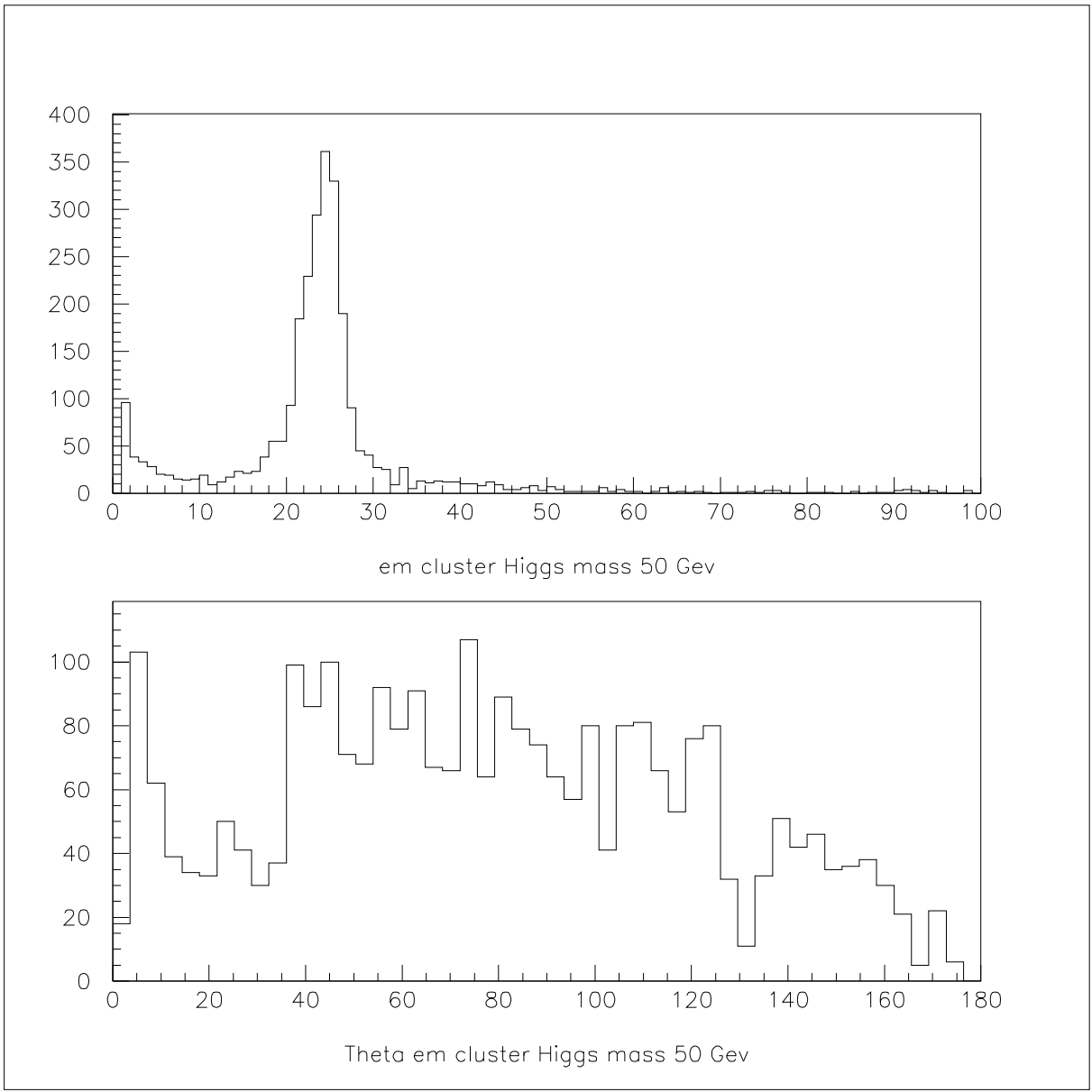

Figure 3: **Energy in the EmCAL**
a) Electron energy in the electromagnetic cluster
b) Θ of electromagnetic cluster of one electron in the final state
both distributions are obtained with $M_{H^{--}} = 50\ GeV$

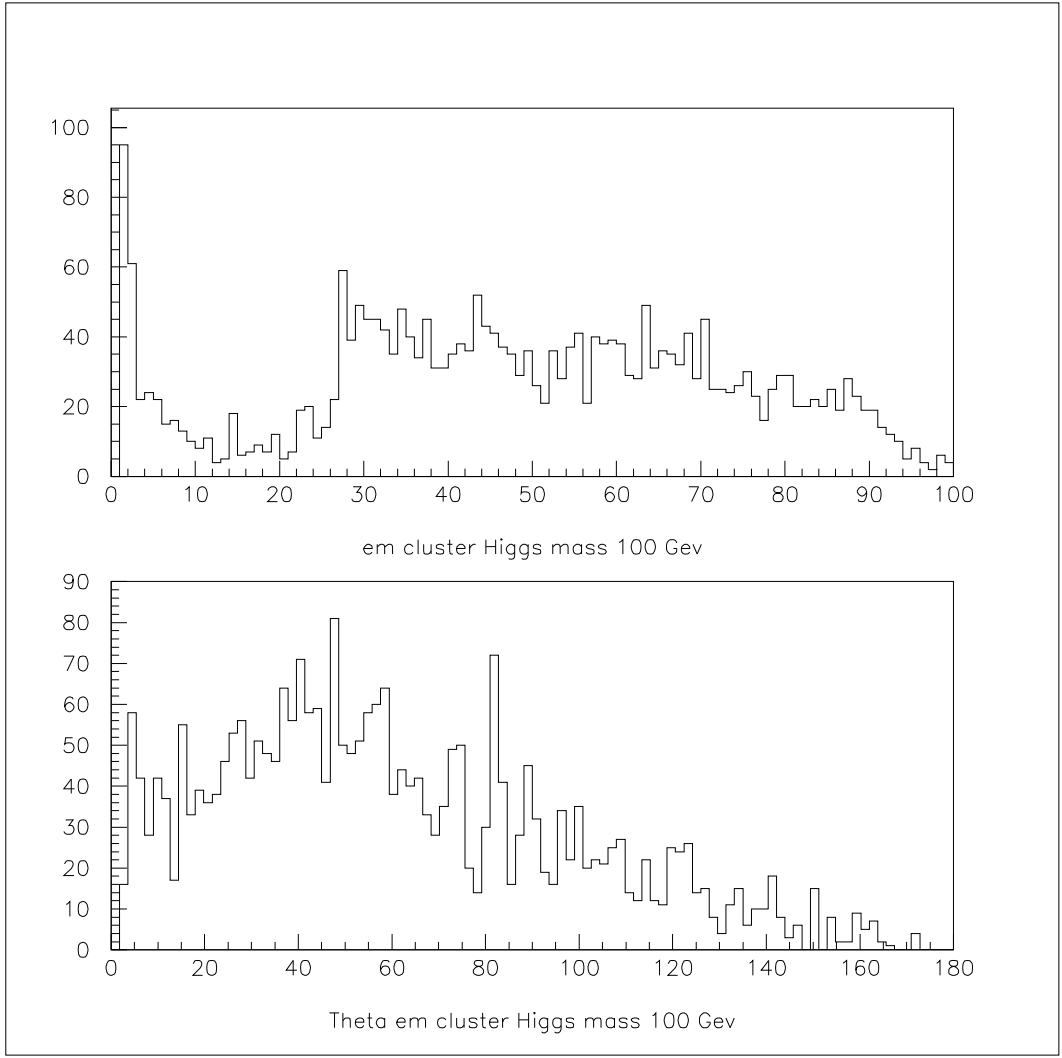

Figure 4: **Energy in the EmCAL**
a) Electron energy in the electromagnetic cluster
b) Θ of electromagnetic cluster of one electron in the final state
both distributions are obtained with $M_{H^{--}} = 100\ GeV$

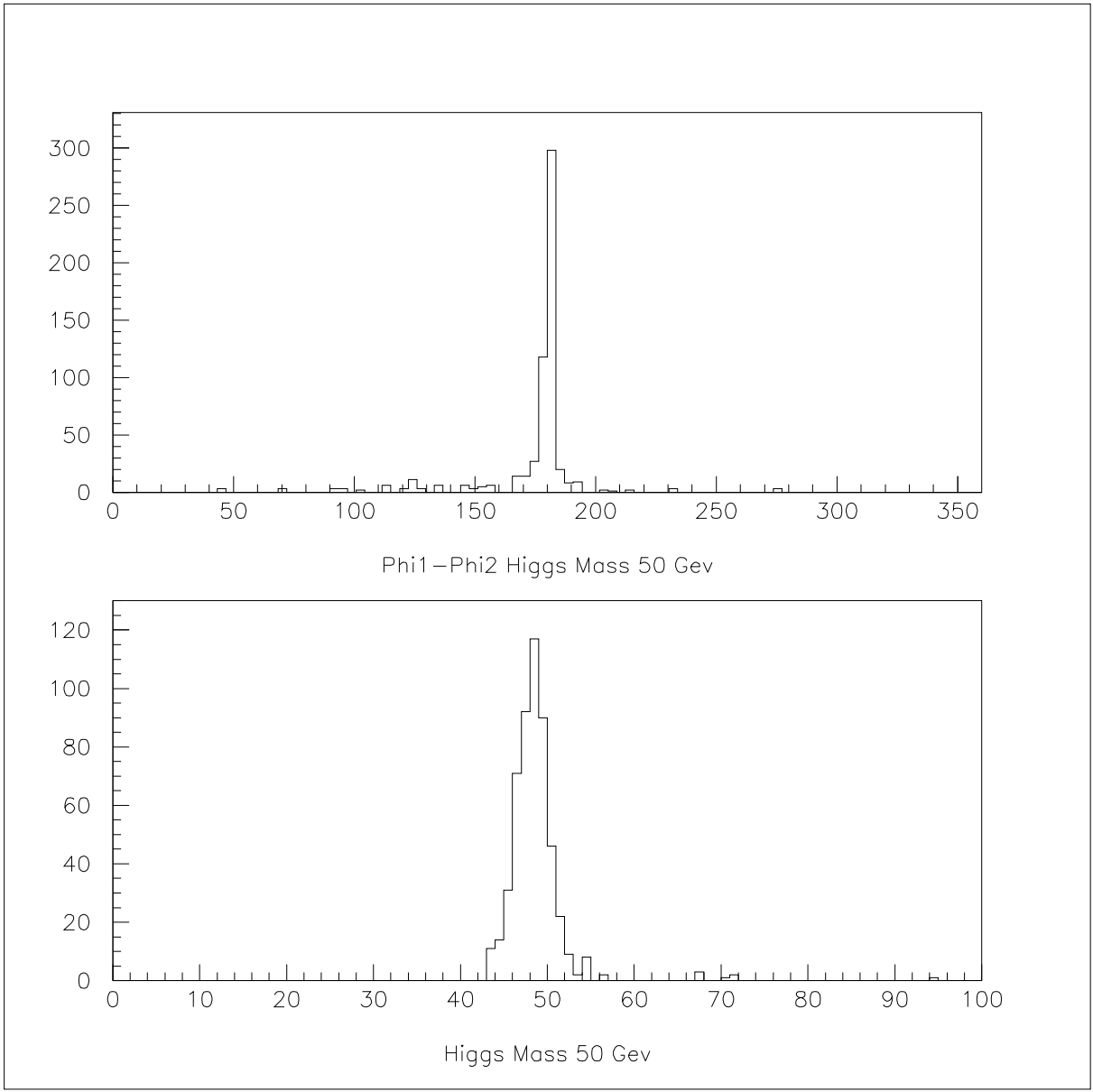

Figure 5: $M_{H^{--}} = 50 \; GeV$:
a) $\Delta\phi_{ee}$ distribution of the electromagnetic cluster for the two highest electron
b) Higgs mass evaluated using the electromagnetic clusters; the cuts are discussed in the text

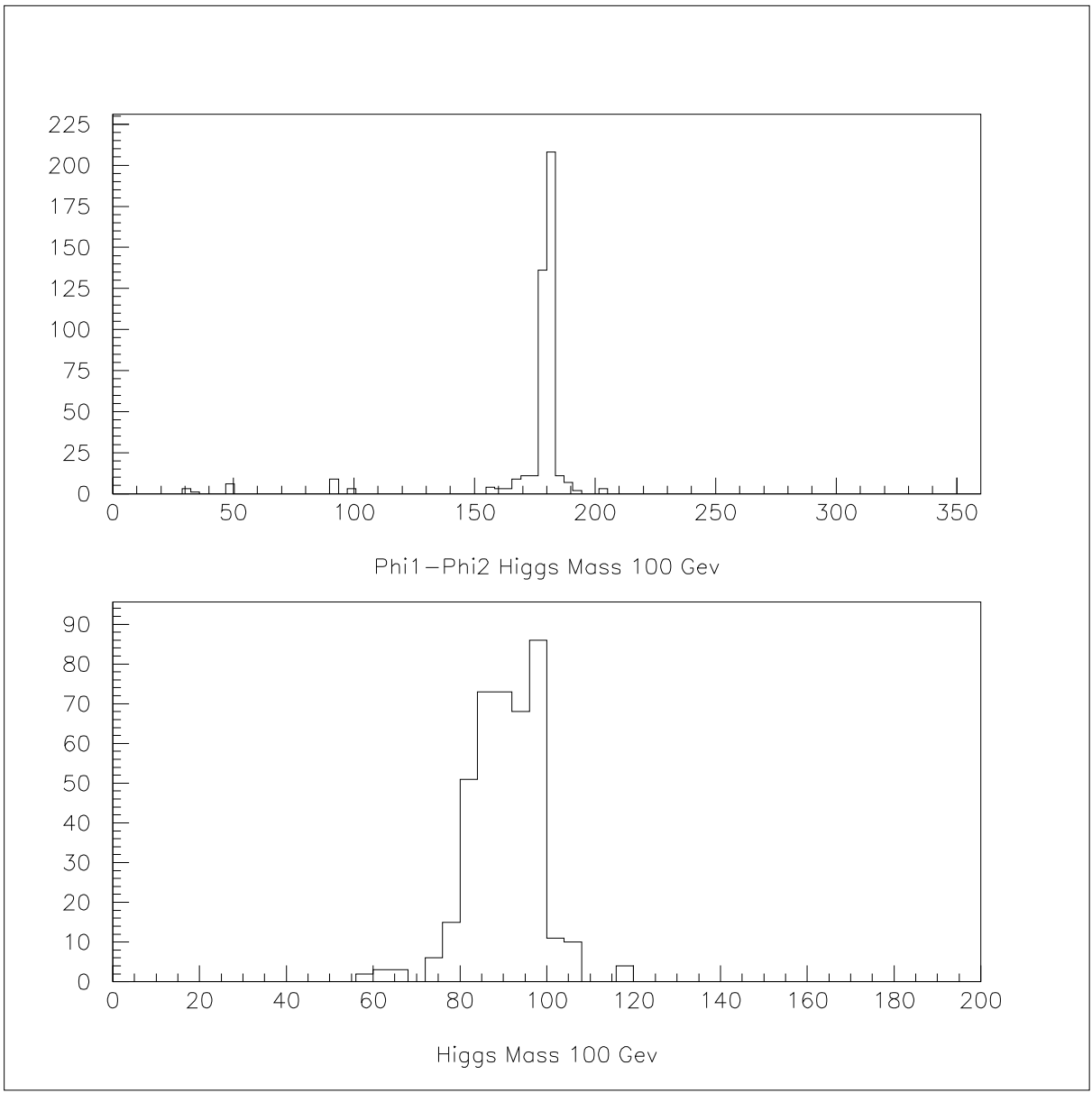

Figure 6: $M_{H^{--}} = 100\ GeV$:
a) $\Delta\phi_{ee}$ distribution of the electromagnetic cluster for the two highest electron
b) Higgs mass evaluated using the electromagnetic clusters; the cuts are discussed in the text

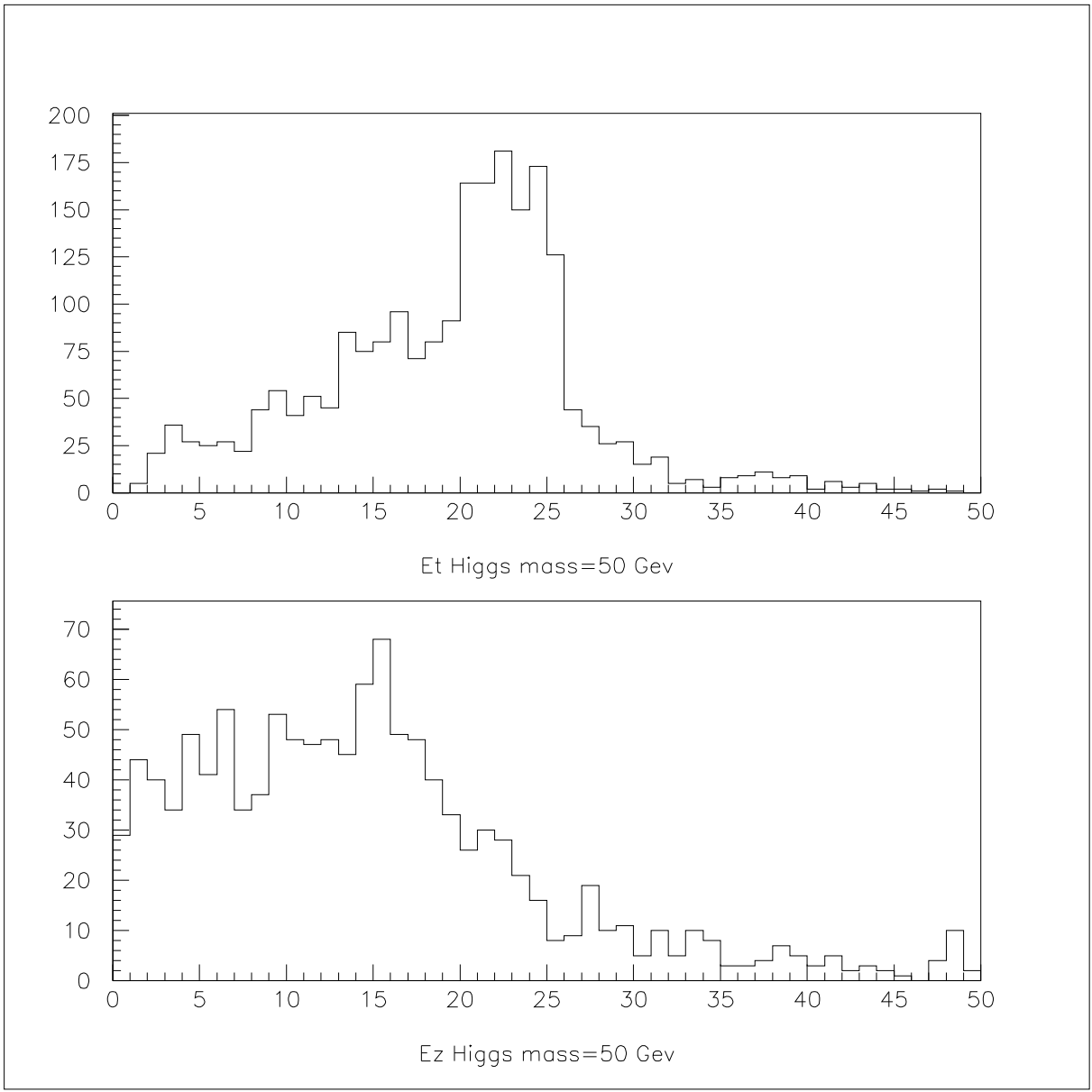

Figure 7: $M_{H^{--}} = 50\ GeV$:
Transverse energy, $E_t$ and longitudinal energy, $E_z$ for the decay electron with $E_e > 20\ GeV$

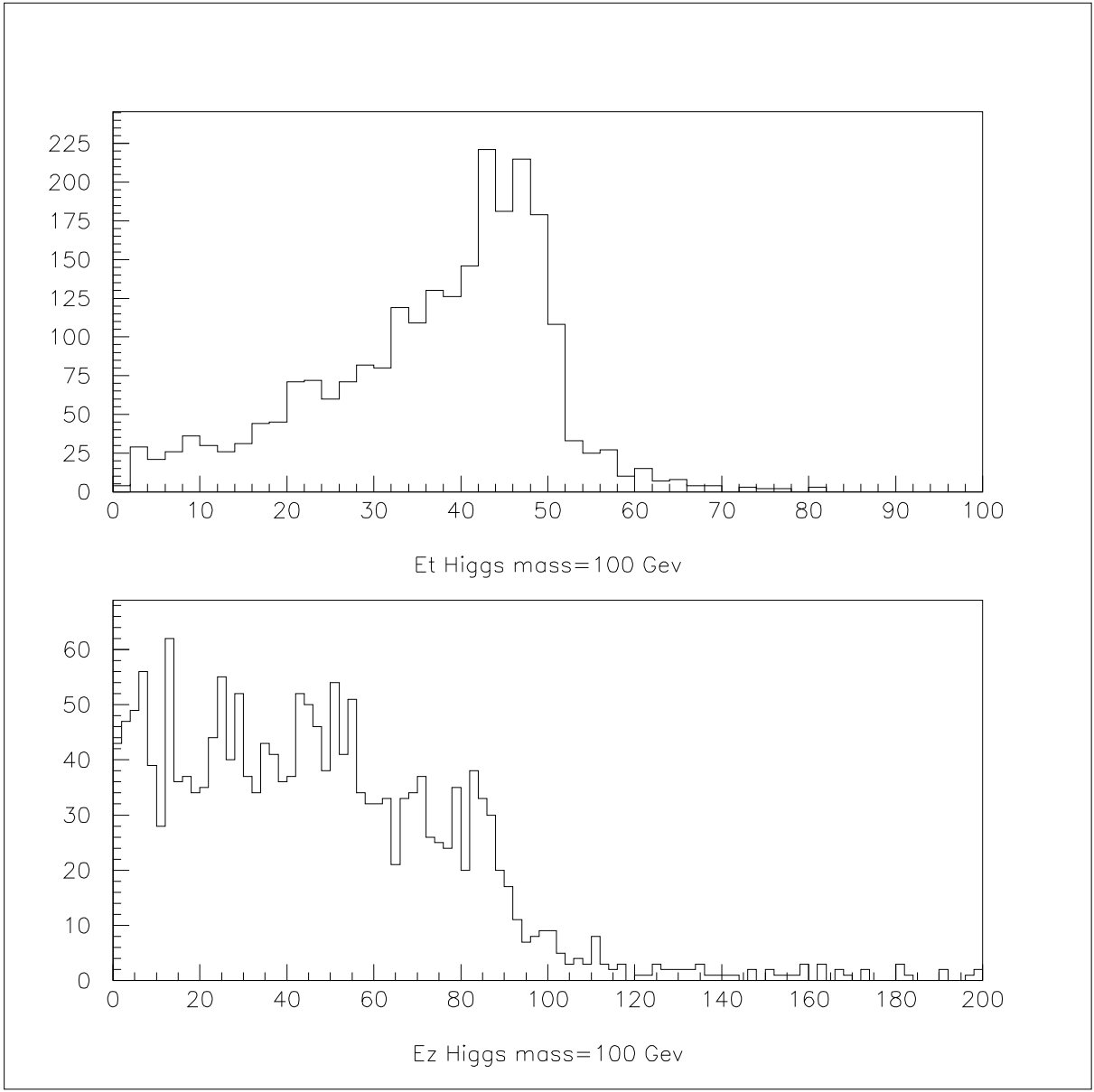

Figure 8: $M_{H^{--}} = 100 \ GeV$:
Transverse energy, $E_t$ and longitudinal energy, $E_z$ for the decay electron with $E_e > 20 \ GeV$